\documentclass{article}
\usepackage{spconf,amsmath,graphicx,array,tablefootnote, url, tikz}

\usepackage{caption, subcaption}

\let\OLDthebibliography\thebibliography
\renewcommand\thebibliography[1]{
  \OLDthebibliography{#1}
  \setlength{\parskip}{0pt}
} 

\newcommand\copyrighttext{%
  \footnotesize 2024 IEEE.  Personal use of this material is permitted.  Permission from IEEE must be obtained for all other uses, in any current or future media, including reprinting/republishing this material for advertising or promotional purposes, creating new collective works, for resale or redistribution to servers or lists, or reuse of any copyrighted component of this work in other works.}
\newcommand\copyrightnote{%
\begin{tikzpicture}[remember picture,overlay]
\node[anchor=south,yshift=10pt] at (current page.south) {\fbox{\parbox{\dimexpr\textwidth-\fboxsep-\fboxrule\relax}{\copyrighttext}}};
\end{tikzpicture}%
}

\title{An Experimental Comparison Of Multi-view Self-supervised Methods For Music Tagging}
\name{Gabriel Meseguer-Brocal, Dorian Desblancs, and Romain Hennequin}
\address{Deezer Research, Paris, France}
\begin{document}
%

\maketitle
\begin{abstract}
Self-supervised learning has emerged as a powerful way to pre-train generalizable machine learning models on large amounts of unlabeled data. 
It is particularly compelling in the music domain, where obtaining labeled data is time-consuming, error-prone, and ambiguous. 
%
During the self-supervised process, models are trained on pretext tasks, with the primary objective of acquiring robust and informative features that can later be fine-tuned for specific downstream tasks. The choice of the pretext task is critical as it guides the model to shape the feature space with meaningful constraints for information encoding.
%
In the context of music, most works have relied on contrastive learning or masking techniques. In this study, we expand the scope of pretext tasks applied to music by investigating and comparing the performance of new self-supervised methods for music tagging.
We open-source a simple ResNet model trained on a diverse catalog of millions of tracks.
Our results demonstrate that, although most of these pre-training methods result in similar downstream results, contrastive learning consistently results in better downstream performance compared to other self-supervised pre-training methods. This holds true in a limited-data downstream context.
\end{abstract}

\copyrightnote

\section{Introduction}
\label{sec:intro}

Over the past few years, self-supervised learning methods have become popular for training robust, generalizable machine learning models \cite{kenton2019bert, baevski2020wav2vec}.
These methods eliminate the need for labeled data. 
Instead, we broadly define self-supervised learning as a paradigm in which a model is trained to accomplish a task whose ground truth is trivially generated from the data itself.
This task, known as  ``pretext", may not have practical interest, but its resolution requires the machine to encode reusable generic signal information efficiently.
These approaches offer numerous advantages. First, they allow the use of extremely large datasets for pre-training and can thus help mitigate some of the biases and ambiguities associated with labeled data~\cite{goyal_2022, hendrycks2019using}. Moreover, self-supervised models can produce features that are not specialized for solving specific supervised tasks. They instead encapsulate richer mid-level representations that make them adaptable to various downstream scenarios with minimal effort and resources; often, a single multilayer perceptron (MLP) trained upon self-supervised embeddings is sufficient for competitive downstream performance \cite{wang2022towards, ericsson2021well}.
Presently, self-supervised learning approaches have achieved competitive or even superior results compared to traditional supervised learning methods. 
They have also paved the way for new research directions, such as learning 
multimodal spaces for text and image \cite{radford_2021} or vision and audio \cite{arandjelovic2017look}.

Self-supervised methods rely on several key elements, such as the choice of data, the method for generating different input views, and the definition of the pretext task. 
The data used plays a critical role in determining the generalization scope of the trained model; generating diverse data views defines the specific signal traits to which the model becomes variant and invariant to; and the pretext task defines the supervised learning objective that is used to update the model.
While all of these aspects are essential in defining a proper self-supervised task, in this study, we focus on understanding the impact of the pretext task on the overall effectiveness of a pre-trained model in the context of music. 

Self-supervised approaches can be clustered into two main categories that are based on the definition of the pretext task \cite{balestriero2023cookbook}.
In the first, the self-supervised method is defined for a single view of the input signal. The model is then trained to predict missing or altered parts of the data, which usually leads to it learning useful properties of the data.
These methods include masking techniques commonly employed by large language models \cite{brown2020language} or spectral inpainting \cite{tagliasacchi2020pre}. 
On the other hand, tasks in the second category rely on several views of the same data point. 
The model is then trained to generate embeddings that are similar when the input consists of different views of the same data instance. 
Data augmentation techniques are often used to create these diverse views. However, the challenge lies in ensuring that the models do not collapse into a trivial constant solution, where everything is deemed similar to everything else, as highlighted in \cite{jing2021understanding}. To address this issue, different methods diverge in how they measure agreement and prevent collapse.
In the realm of music, contrastive learning, a popular pretext task in this category, which we will outline later in this paper, has proved to be effective for numerous tasks, such as tagging \cite{mccallum_2022, spijkervet2021contrastive}, classification \cite{saeed2021contrastive}, and even beat tracking \cite{desblancs2023zero} and artist identification \cite{yakura2022self}.
However, more recent pretext tasks that have gained traction in the computer vision community are yet to be benchmarked on popular music information retrieval tasks.

Hence, this paper focuses on the second category of pretext tasks applied to classic music tagging tasks.
We pre-train a simple ResNet architecture \cite{he2016deep} on the same, large-scale music dataset with five popular pretext tasks: contrastive learning \cite{chen2020simple}, Bootstrap your own Latent (BYOL) \cite{grill_2020}, clustering \cite{caron_2020, caron_2021}, Barlow twins \cite{zbontar_2021}, and Variance-Invariance-Covariance Regularization (VICReg) \cite{bardes_2022, bardes_2022_local}.
The embeddings generated by each model are then used to train a single-layer MLP model for five downstream tagging tasks.
We also report performance on these downstream tasks in a limited-data setting, where only small portions of the training set are used.
Our results demonstrate that the models trained upon contrastive learning embeddings consistently outperform the others, though the gap is quite narrow.
We hope that these findings will aid researchers and engineers in the music or audio industry in selecting the best-performing pre-trained model for their needs.
Furthermore, we open-source the trained models,\footnote{ \scriptsize\url{https://github.com/deezer/multi-view-ssl-benchmark}} with the hope that the same people will be able to take advantage of the scale of the catalog we used for training, and perhaps investigate the musical features that the generated embeddings encode.

\section{Pretext Methodology}\label{sec:methodology}

\begin{table}[t]
\centering
\resizebox{.8\columnwidth}{!}{%
{\renewcommand{\arraystretch}{1.1}%
\begin{tabular}{l c c c c c}
\text{Dataset} &  \text{\texttt{\#}Labels} & \text{\texttt{\#}Train} & \text{\texttt{\#}Val.} & \text{\texttt{\#}Test} & Ref. \\ \hline
MSD100  & 100 & 71388 & 15618 & 15281 & \cite{bertin2011million}\tablefootnote{Splits from: \tiny \url{https://github.com/minzwon/tag-based-music-retrieval}}\\
MTAT & 50 & 15244 & 1529 & 4332 & \cite{law2009evaluation}\textsuperscript{3}\\
Jam\textsubscript{Top50} & 50 & 32136 & 10888 & 11356 & \cite{bogdanov2019mtg}\tablefootnote{We use the first provided split for each Jamendo experiment.} \\
Jam\textsubscript{Instrument} & 40 & 14395 & 5466 & 5115 & \cite{bogdanov2019mtg}\textsuperscript{4} \\
Jam\textsubscript{Mood} & 56 & 9949 & 3802 & 4231 & \cite{bogdanov2019mtg}\textsuperscript{4}
\end{tabular}}%
}
\caption{Overview of the downstream datasets.}
\label{tab:down}
\vspace*{-.3cm}
\end{table}

This study compares the impact of self-supervised pretext tasks that leverage multiple views of the same data point. 
Our methodology is, therefore, consistent across the pretraining pipeline. 
We notably ensure that the data, view generation, and model architecture remain constant throughout all approaches in order to better measure the impact of the pretext definition on downstream performance. 

 \begin{figure}[t]
     \centering   
     \includegraphics[width=.91\columnwidth]{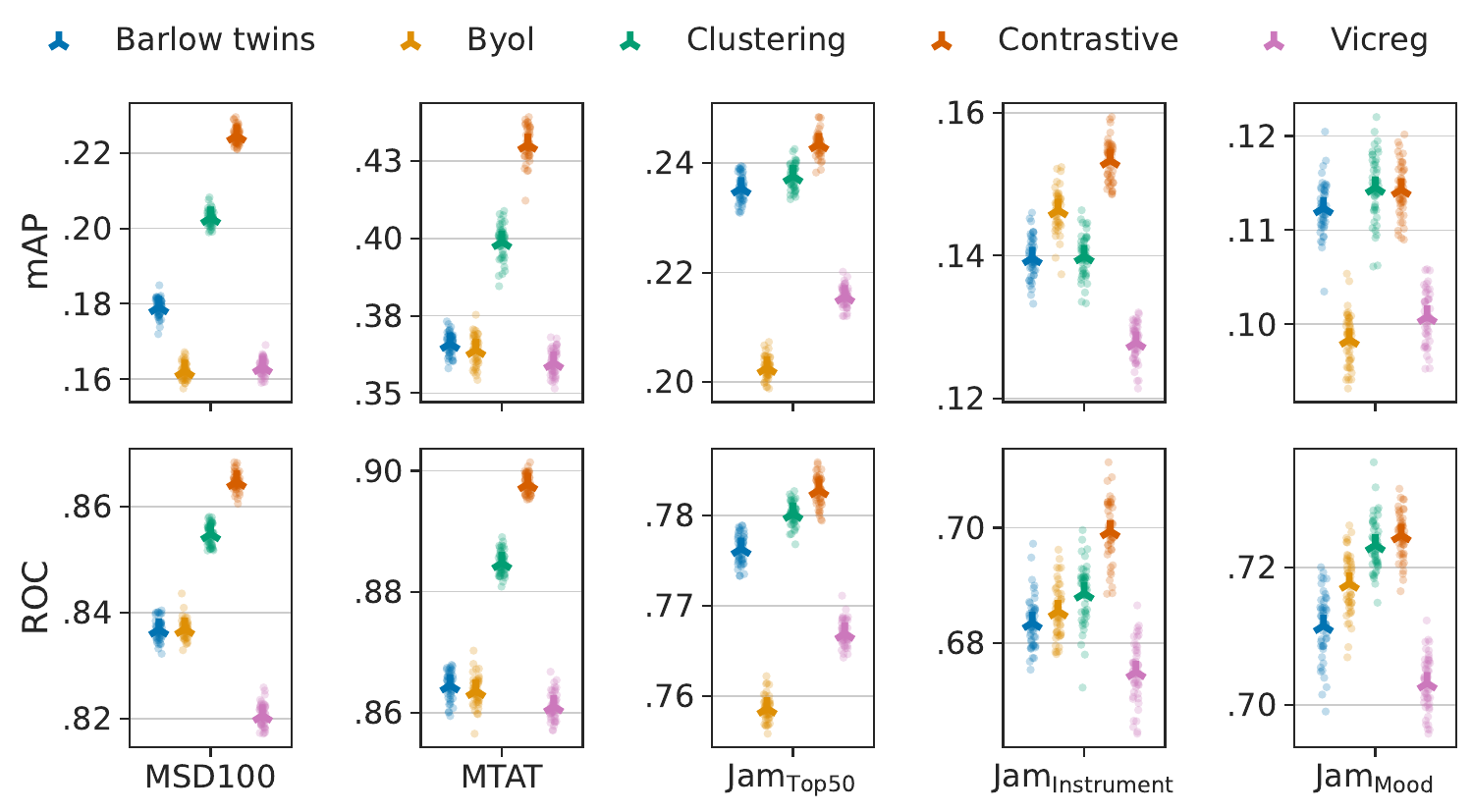}
     \caption{Downstream results. We apply transfer learning to each task by training an MLP classifier on top of the embeddings generated by the frozen pretext model. We utilize bootstrapping. Each dot represents the metric of a resampled batch. The marker indicates the mean of each result.
     }
     \label{fig:exp1}
 \end{figure}

\subsection{Data}\label{sec:data-train}

We use an in-house dataset, which consists of $\sim$4M full tracks, to pre-train our models. 
We aim to make our dataset as musically balanced and diverse as possible in order to expand the models' generalization ability toward unseen data.
To do so, we collect tracks that have been used in past curated playlists from the music streaming service Deezer.
These playlists are categorized by mood or genre; our dataset thus contains a vast collection of high-quality sounds and songs that span various genres and periods. 
We acknowledge that these may be biased toward Western and Billboard music.
Each piece of audio is resampled at 16 kHz, normalized, and converted to mono.
We then compute a log-magnitude mel-spectrogram with 128 frequency dimensions to use as input for each of our models. 

In music, multi-view methods rely on the idea that various facets of music, such as songs, albums, or artists, collectively share common features \cite{alonso_jimnez_2022}, which are sufficiently rich and informative to extract meaningful music-invariant traits. 
For each self-supervised approach, the input data consists of two four-second audio pairs, which we call \textit{anchor} and \textit{positive}.
For each song, an anchor is randomly selected. 
The corresponding positive is then selected from the same song. 
We force this segment to be at least four seconds away from the anchor but no further than sixteen seconds away in order to ensure both segments are contextually similar. 
Furthermore, we use online training: every time we see an audio track, the constructed pairs are generated on the fly.
All self-supervised models are then trained with batches consisting of 256 pairs for 1000 epochs of 512 steps.

\subsection{Architecture}
\label{sec:architecture}

Our self-supervised learning models have a ResNet backbone architecture and contain a total of 2.8 million trainable parameters. 
Similarly to \cite{manco2022}, we incorporate an attention layer that summarises the sequential data into a single, 1024-dimension embedding vector. 
A projector head is then placed upon the backbone to solve our designated pretext task. 
This ensures that our generated embeddings are not tailored to the pre-training method; they maintain a broader, more general utility.
All the approaches we study in this work use the same head.
It is comprised of two blocks, which consist of a linear layer, batch normalization, and a ReLu activation.
The first linear layer contains 1024 connected weights, while the final linear 2048.
All our approaches are trained using a single GPU.\footnote{NVIDIA GeForce GTX TITAN X}

 \begin{figure}[t]
     \centering
     \includegraphics[width=.91\columnwidth]{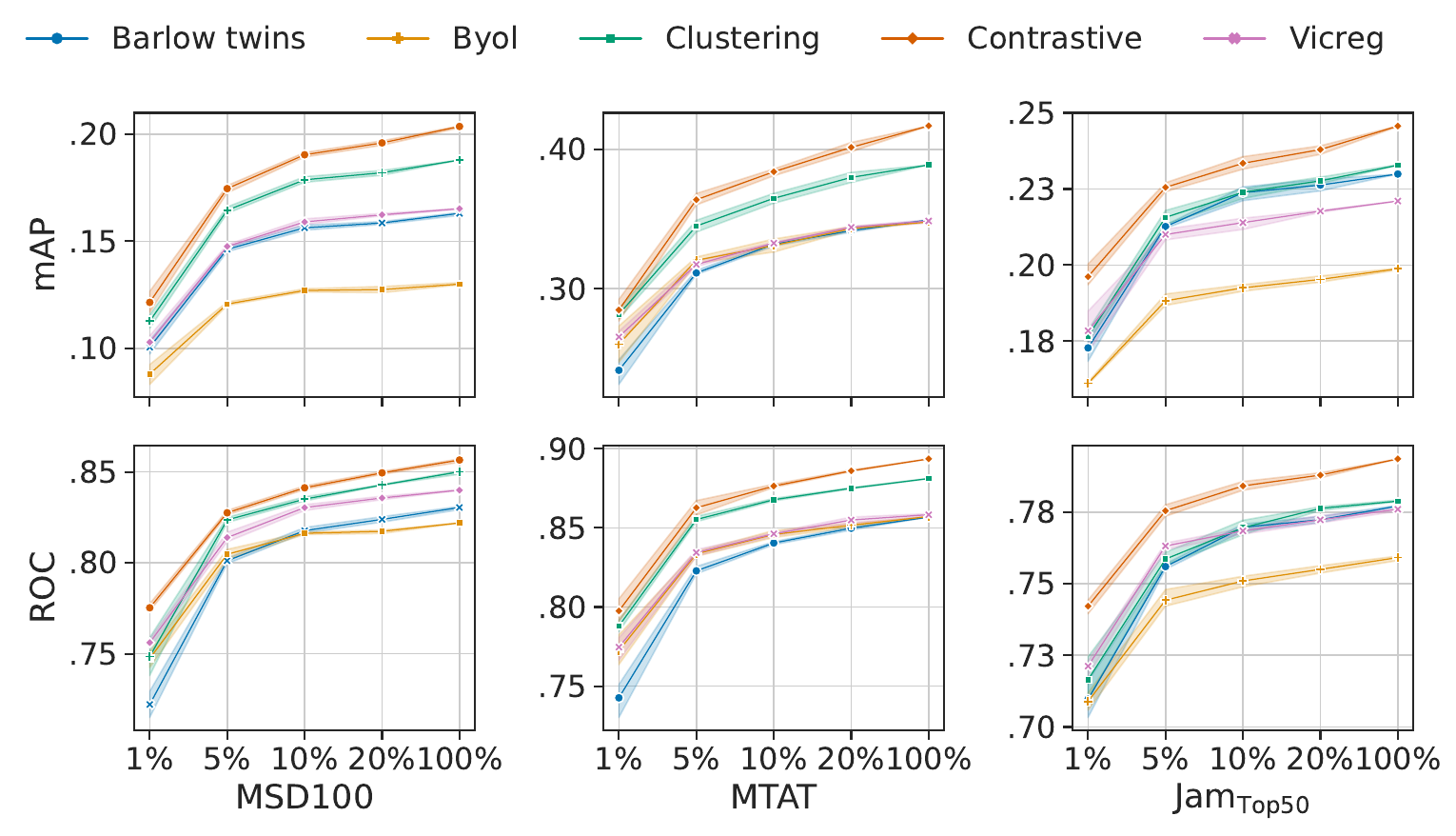}
     \caption{Limited data results. Each dataset's train set is randomly sampled four times at four different percentages. We report the mean test set metrics obtained for each approach. We also include the results using the full train set in order to showcase each model's performance in comparison to these.
     }
     \label{fig:exp2}
 \end{figure}

\subsection{Pre-training Tasks}
\label{sec:approaches}
Our paper studies how five different self-supervised pretext tasks popular in image processing adapt to the audio domain. 

\textbf{Contrastive} learning aims to project similar samples closer together in the embedding space while pushing dissimilar samples apart. 
We use the normalized temperature-scaled cross-entropy loss \cite{sohn_2016}, with a temperature set to 0.1, in order to do so. 
In this setting, similarly to \cite{spijkervet2021contrastive}, for each anchor and positive pair, negative examples consist of the rest of the audio segments in the same batch. 
The distance between embeddings is computed using a cosine similarity.

\textbf{BYOL} \cite{grill_2020} uses a teacher-student approach. 
The student model is encouraged to match a target embedding representation generated by the teacher. 
The networks are almost identical and start with the same random initialization.
The student, however, needs an extra prediction layer for stability reasons.
The distance used to compare embeddings is a mean squared error between the normalized student predictions and target teacher projections.
At each training step, the student is updated with the normal gradient, whereas the teacher is updated using an exponential moving average (EMA) of recent student weights. 
Unlike in this work, in \cite{niizumi2021byol}, a unique, data-augmented segment of audio is used for BYOL pre-training.
We hope to explore whether this type of pre-processing leads to more favorable self-supervised learning in the future.

\textbf{Clustering} \cite{caron_2020, caron_2021} groups embeddings based on similarity. 
It uses a teacher-student approach close to BYOL. 
However, instead of computing similarity between embeddings, the teacher generates the target class distribution the student has to match via a cross-entropy loss. 
Both models are initialized identically and have the same architecture.
At each training step, the student is updated normally, whereas the teacher is updated with an EMA. 
As presented in \cite{caron_2020}, we use centering and sharping to avoid collapsing into a single class and uniform distribution.

The two methods are grounded in the imposition of statistical properties within the embedding space. 
\textbf{Barlow-Twins} \cite{zbontar_2021} encourages the diagonalization and independence of each embedding dimension.
The pretext task's goal is to ensure that the cross-correlation between the anchor and positive embeddings is close to the identity matrix.
This reduces the redundancy between the components of each embedding dimension.
\textbf{VICReg} \cite{bardes_2022, bardes_2022_local} combines variance, invariance, and covariance terms. 
The invariance term encourages proximity between the anchor and positive embedding vectors by minimizing their mean squared distance. 
The variance term promotes diversity among embedding vectors within a batch. 
Finally, the covariance term is similar to Barlow-Twins, but it focuses exclusively on reducing off-diagonal terms, effectively decorrelating the variables within each embedding.

\section{Evaluation}
\label{sec:downstream}

\subsection{Downstream Tasks}\label{sec:downstream}
We apply the same procedure across all the downstream tasks and pretext approaches. 
We use an MLP that operates directly on the average embedding space of each track to find the linear separation between classes. Thus, we optimize $1024 \times N$ weights, where $N$ denotes the number of classes for each downstream class. The backbone generating the embeddings remains frozen. We evaluate the effectiveness and generalizability of each pretext approach on a set of five music tagging datasets, which are depicted in Table \ref{tab:down}. 
We use the standard evaluation metrics: area under the Receiver Operating Characteristic curve (ROC) and mean average precision (mAP).

\subsection{Pretext to Downstream Transfer Learning} 
\label{sec:exp1}
All downstream tasks are first trained using their full training set. 
We use a batch size of 256 pairs and 25 epochs of 128 steps. 
We use the last checkpoint of the pretext backbone and the best downstream model checkpoint, in terms of validation loss, for test set evaluation. 
Both ROC and mAP are computed using all song classifications for the test set; therefore they do not provide insights into the metric values' robustness to data variation. In order to overcome this limitation, we use bootstrapping.
The test dataset is sampled without replacement several times and computes the mAP and ROC for each batch.
We sample $50\%$ of each test set 50 times and report the mean and standard deviation of all samples.

Results in Figure ~\ref{fig:exp1} reveal that 
the model pre-trained using contrastive learning consistently achieves superior performance compared to those pre-trained with other pretext tasks on both metrics. This observation is noteworthy, particularly since this is not the case in the image domain, which serves as inspiration for the majority of this work.
Furthermore, despite other works using contrastive learning with more advanced architectures and incorporating more intricate probes~\cite{mccallum_2022, li2023mert}, our simple ResNet and MLP combination delivers comparable performance. 
In second place, clustering exhibits strong performance, most likely due to its ability to create well-defined groupings within the embedding space.
We observe an uncharted collapse mode, where the model utilizes only a subset of clusters to partition the feature space, preventing the use of this pretext task to its full potential.
On the contrary, the BYOL method exhibits lower performance, which can be attributed to its distinct design. 
Predicting the same embedding space appears to be overly restrictive and stringent in this context. 
The differing performances of VICReg and Barlow Twins are intriguing. While the simple orthogonalization and diagonalization of Barlow Twins is successful for conventional tagging tasks like MSD100 and Jam\textsubscript{Top50}, it proves insufficient for separating instruments and mood, where VICReg's regularization is better suited.

\subsection{Limited Data Music Tagging} 
\label{sec:exp2}
Annotating a music dataset is a time-consuming, error-prone process frequently plagued by inherent ambiguities \cite{bittner2021vocadito}. As a result, a substantial portion of datasets within the MIR community remain relatively small in scale, which hinders our ability to train deep neural networks on them. In this experiment, our objective is to assess the adaptability of each approach when faced with the constraint of limited data availability. 
We focus on three datasets: MSD100, MTAT, and Jam\textsubscript{Top50}. 
We randomly sample each dataset's train set four times for four different percentages ($1\%$, $5\%$, $10\%$, and $20\%$) and train a different MLP for each iteration and percentage using the same specifications as earlier. We then evaluate their performance on the full test set, as shown in Figure \ref{fig:exp2}.
It is interesting to note that all approaches demonstrate decent performance with just $1\%$ of the training set and nearly optimal results with just $10\%$. Subsequent performance improvements occur at a slower rate.
We also observe that, as seen previously, contrastive learning outperforms all other methods. It is, however, worth mentioning that the performance gap between clustering and contrastive methods is less pronounced when using limited data compared to the full dataset. Nonetheless, all other pre-training methods perform similarly and less well than the two methods mentioned previously.

\subsection{Training Stability} \label{sec:stability}
In our empirical findings, we observed that contrastive learning and Barlow Twins are the most stable methods during training. They both avoid convergence issues. These models also rely on a minimal set of hyperparameters, 
the scale parameter for the former and diagonal and off-diagonal ratios for the latter.
On the other hand, VICReg is highly sensitive to hyperparameter choice since it uses multiple losses.
This may explain its notable fluctuations in performance.
BYOL exhibits instability. We observe substantial fluctuations in both training and validation loss from one epoch to another. It is highly sensitive to the EMA momentum, where slight value variations have a significant impact on the model weights. It is also worth noting that it requires the most time per backpropagation step compared to other approaches.
Clustering demonstrates an even greater reliance on hyperparameter selection, as it encompasses a complex interplay of factors, such as the EMA momentum for teacher and centering updates, temperature settings for the sharping and centering process to avoid collapse and weight decay. Striking the right balance among these elements is delicate and often requires the use of multiple schedulers to achieve effective model optimization.
We found that the default values for most hyperparameters proposed in the original works for image processing are not well-suited for audio applications. As a result, we introduced new parameter values to mitigate early training plateaus. It is worth noting that further hyperparameter tuning could lead to improved results, a direction that warrants exploration in future investigations. Our settings can all be visualized in the GitHub repository linked with this publication.

\section{Conclusion}
\label{sec:conclusion}

This paper presents a comparative analysis of various self-supervised pretext tasks for music tagging: contrastive learning, BYOL, Barlow Twins, VICReg, and clustering. 
A simple ResNet model is pre-trained with these methods. From there, an MLP is trained upon the embeddings generated by these methods to solve the downstream tagging tasks we selected in this work.
Our work highlights the importance of choosing a relevant pretext task that aligns with our specific domain. 
For our downstream tasks, contrastive learning stands out as the preferred choice. It consistently demonstrates superior performance and relies minimally on hyperparameter tuning. 
Clustering shows promise. However, its performance is strongly linked to hyperparameter tuning and may be affected by the uncharted collapse we observed.
Strategies to address this issue are to be addressed in future research.
Finally, models pre-trained with BYOL, Barlow Twins, and VICReg do not perform as well as models trained with the two pretext tasks mentioned earlier in this paragraph, even in a limited-data setting.
Our models and code are readily accessible; 
we hope they can be a valuable resource to members of our community who would like to take advantage of the training scale used or investigate the musical features the generated embeddings encode.
\vfill\pagebreak



\newpage
\bibliographystyle{IEEEbib}
{\small
\bibliography{Template}}

\end{document}